\documentclass[aps,reprint, prl, superscriptaddress, showpacs,nobibnotes]{revtex4-1}
\usepackage{graphicx}
\usepackage{filecontents}
\usepackage{xcolor}

\begin{document}

\title{Sudden collapse of magnetic order in oxygen deficient nickelate films}%

\author{Jiarui Li}%
\affiliation{Department of Physics, Massachusetts Institute of Technology, Cambridge, Massachusetts 02139, USA}

\author{Robert J. Green}%
\affiliation{Department of Physics and Engineering Physics, University of Saskatchewan, Saskatoon, Saskatchewan, Canada S7N 5E2}
\affiliation{Stewart Blusson Quantum Matter Institute, University of British Columbia, Vancouver, British Columbia, Canada V6T 1Z4}

\author{Zhen Zhang}%
\affiliation{School of Materials Engineering, Purdue University, West Lafayette, IN 47907}

\author{Ronny Sutarto}%
\affiliation{Canadian Light Source, Saskatoon, SK S7N 2V3, Canada}

\author{Jerzy T. Sadowski}%
\affiliation{Center for Functional Nanomaterials, Brookhaven National 	Laboratory, Upton, NY 11973}

\author{Zhihai Zhu}%
\affiliation{Department of Physics, Massachusetts Institute of Technology, Cambridge, Massachusetts 02139, USA}

\author{Grace Zhang}%
\affiliation{Department of Physics, Massachusetts Institute of Technology, Cambridge, Massachusetts 02139, USA}

\author{Da Zhou}%
\affiliation{Department of Physics, Massachusetts Institute of Technology, Cambridge, Massachusetts 02139, USA}

\author{Yifei Sun}%
\affiliation{School of Materials Engineering, Purdue University, West Lafayette, IN 47907}

\author{Feizhou He}%
\affiliation{Canadian Light Source, Saskatoon, SK S7N 2V3, Canada}

\author{Shriram Ramanathan}%
\affiliation{School of Materials Engineering, Purdue University, West Lafayette, IN 47907}

\author{Riccardo Comin}%
\email{rcomin@mit.edu}
\affiliation{Department of Physics, Massachusetts Institute of Technology, Cambridge, Massachusetts 02139, USA}

\begin{abstract}
Oxygen vacancies play a crucial role in the control of the electronic, magnetic, ionic, and transport properties of functional oxide perovskites. Rare earth nickelates (RENiO$_{3-x}$) have emerged over the years as a rich platform to study the interplay between the lattice, the electronic structure, and ordered magnetism. In this study, we investigate the evolution of the electronic and magnetic structure in thin films of RENiO$_{3-x}$, using a combination of X-ray absorption spectroscopy and imaging, resonant X-ray scattering, and extended multiplet ligand field theory modeling. We find that oxygen vacancies modify the electronic configuration within the Ni-O orbital manifolds, leading to a dramatic evolution of long-range electronic transport pathways despite the absence of nanoscale phase separation. Remarkably, magnetism is robust to substantial levels of carrier doping, and only a moderate weakening of the $(1/4, 1/4, 1/4)_{pc}$ antiferromagnetic order parameter is observed, whereas the magnetic transition temperature is largely unchanged. Only at a certain point long-range magnetism is abruptly erased without an accompanying structural transition. We propose the progressive disruption of the 3D magnetic superexchange pathways upon introduction of point defects as the mechanism behind the sudden collapse of magnetic order in oxygen-deficient nickelates. Our work demonstrates that, unlike most other oxides, ordered magnetism in RENiO$_{3-x}$ is mostly insensitive to carrier doping. The sudden collapse of ordered magnetism upon oxygen removal may provide a new mechanism for solid-state magneto-ionic switching and new applications in antiferromagnetic spintronics.
\end{abstract}

\date{\today}%
\maketitle


Perovskite-type $3d$ transition-metal oxides (TMOs) realize many interesting electronic phenomena due to their flexibility in accommodating ionic species of varying size and their tolerance to off-stoichiometric chemical compositions. The phase diagrams of these systems host a multitude of broken-symmetry electronic phases, which are often coexisting and intertwined \cite{Fradkin2015,Dagotto2005,Keimer2015}. In TMOs, oxygen vacancies, whether naturally formed or artificially introduced, provide a very effective avenue to alter the electronic properties of TMOs and in turn suppress, enhance, or engender new emergent phases of matter \cite{Tallon, Keimer2015, Herranz2007, Croft1999, Sanchez1996}. Oxygen vacancies have been shown to enhance a broad array of functional properties, including ionic conductivity \cite{Gouget2019}, photoluminescence \cite{Kan2005,Mochizuki2005}, thermal resistance \cite{Chen2018}, electrical conductivity\cite{Herranz2007}, and even superconductivity \cite{Li2019,Tallon}. Although the role of oxygen vacancies differs from system to system, one would expect that, in a mean field scenario, the removal of oxygen atoms changes the electronic configuration and (potentially) band filling via charge compensation \cite{Sanchez1996,  Hoffmann2015, Li2013, Seikh2008, Ramezanipour2012}. This uniform modification of carrier density can also alter the magnetic properties \cite{Balamurugan2006}, and long range magnetism is particularly fragile to carrier doping in systems with strong correlations \cite{Keimer2015,Basov2011}. From another perspective, oxygen removal creates local defects that can disrupt long range electronic and magnetic order \cite{Stolen2006}. However, this second scenario has not been observed so far and is less understood.

\begin{figure*}
	\centering
	\includegraphics[width = 2 \columnwidth]{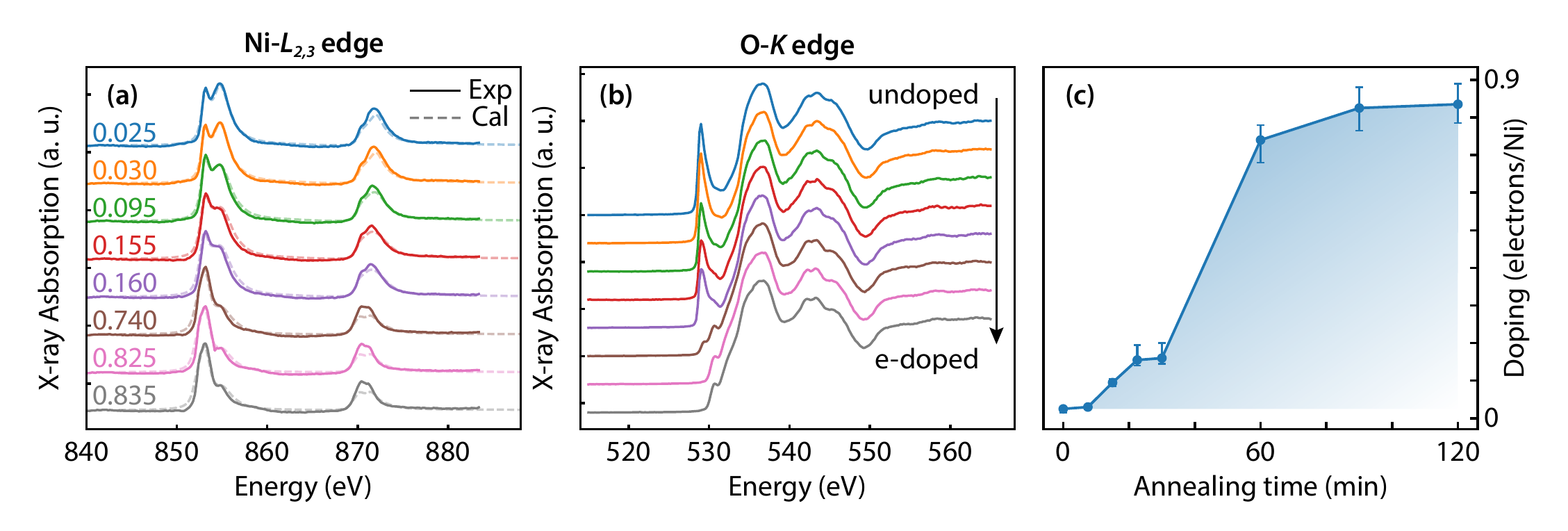}
	\caption{\label{fig1} 
	(a) Measured (solid) and simulated (dashed) X-ray absorption spectra across the Ni-$L_{2,3}$ edges for SmNiO$_{3-x}$ with different electron doping levels, obtained by annealing stoichiometric films in an oxygen deficient atmosphere. The doping level (number of doped electrons per Ni) is specified for each simulated spectrum. Spectra are vertically offset for clarity. 
	(b) X-ray absorption spectra across the O-$K$ edges for the same SmNiO$_{3-x}$ samples. The same color legend is used as the data series in Fig. 1(a).
	(c) The doping levels vs. annealing time for each sample, as extracted from the comparison between experimental and theoretical spectra. Vertical error bars reflect the uncertainty in the doping level for each sample.}
\end{figure*}

The family of rare earth nickelates (RENiO$_3$, where RE = Nd, Sm in this paper) offers new opportunities to study the impact of oxygen removal on the electronic and magnetic properties in a materials system characterized by a complex interplay between the spin, charge, and lattice degrees of freedom. RENiO$_3$ manifest a rich temperature-composition phase diagram which remains the subject of a very active line of inquiry \cite{Catalan2008,Middey2016,Catalano2018}. A pronounced metal-to-insulator transition is followed or accompanied by an antiferromagnetic (AFM) transition \cite{Torrance1992}. The electronic ground state of stoichiometric RENiO$_3$ is characterized as a negative charge-transfer insulator \cite{Zaanen1985}. Here, the oxygen ligand $2p$ electron is self-doped into the Ni $3d$ orbital leading to a ground state with local electronic configuration ${3d}^{7+\delta}$\underline{L}$^\delta$, where \underline{L} denotes an oxygen $2p$ hole \cite{Mizokawa1999,Bisogni2016}. Upon cooling into the insulating phase, a charge/bond disproportionation transition takes place: the NiO$_6$ octahedra undergo a static, long-range breathing distortion with propagation vector $(1/2, 1/2, 1/2)_{pc}$, creating two inequivalent Ni sites. As a result, the electronic degeneracy is further split as:
\begin{equation} \label{eq1}
2 \times {3d}^{7+\delta} \underline{\mathrm{L}}^\delta   \rightarrow 
{3d}^{7+\delta+n_1} \underline{\mathrm{L}}^{\delta+n_2} + 
{3d}^{7+\delta-n_1} \underline{\mathrm{L}}^{\delta-n_2}
\end{equation}
where $n_1,n_2$ represent the magnitude of the charge/bond disproportionation, respectively. Within the bond-disproportionated phase, magnetic order sets in with a supercell composed of 4 lattice units along the body diagonal direction of the pseudocubic unit cell and corresponding propagation vector $(1/4, 1/4, 1/4)_{pc}$. Previous studies have found that the spin texture in the AFM phase is either collinear "up-up-down-down" or non-collinear "up-right-down-left" \cite{Gaecia-Munoz1992,Scagnoli2006,Frano2013,Hepting2018}.

Recently, reversible tuning of the oxygen stoichiometry in thin film samples of RENiO$_3$ has been achieved by means of highly-controlled post annealing procedures \cite{Wang2016,Kotiuga2019}. The creation of oxygen vacancies alters the electronic structure via charge compensation, driving the material into a highly insulating state. Oxygen deficiency thus offers a powerful route to tune the electronic and magnetic ground state of RENiO$_{3-x}$, enabling access to their broader electronic phase diagram \cite{Nikulin2004,Tiwari1999,Garcia-Munoz1995}. Moreover, recent work on oxygen-reduced Nd$_{0.8}$Sr$_{0.8}$NiO$_{2}$ has led to the discovery of superconductivity in the nickelate family \cite{Li2019}, underscoring the importance of studying the ground state properties of oxygen-deficient nickelates.

In the present study, we examine the evolution of electronic and magnetic ground state in RENiO$_{3-x}$ using a combination of extended multiplet ligand field theory, X-ray spectroscopy and resonant soft X-ray scattering. We chart out the electronic and magnetic phase diagram as a function of temperature and oxygen stoichiometry, which reveals the dual role of oxygen vacancies as (electronic) dopants and (magnetic) defects. On the one hand, we find that the removal of oxygens from stoichiometric RENiO$_{3}$ homogeneously injects electrons into the Ni $3d$ and O $2p$ conduction bands. On the other hand, we observe an unusual evolution of $(1/4, 1/4, 1/4)_{pc}$ magnetic order, which is progressively weakened upon oxygen removal but without a significant change in $T_{AFM}$, until it collapses at a doping level of $\sim$ 0.21 e$^{-}$/Ni. The absence of nanoscale spatial inhomogeneity in the electronic ground state upon doping suggests that the collapse of magnetic order is due to the progressive disruption of the superexchange interaction network caused by the random formation of localized oxygen defect sites with removed O $2p$ ligand orbitals.

To understand how the electronic state in RENiO$_{3-x}$ evolves upon doping, we performed X-ray absorption spectroscopy (XAS) measurements on thin films of SmNiO$_{3-x}$ (SNO) and NdNiO$_{3-x}$ (NNO). More details about the sample and experiment can be found in the supplementary material \cite{Sup}. Figure 1 displays the SNO XAS profiles across the Ni $L_{2,3}$ and O $K$ edges at 22 K, the lowest temperature measured in the present study. At this temperature, both undoped SNO and NNO are well within the insulating state, as signaled by the double peak structure at the Ni-$L_3$ resonance (853.2 and 854.8 eV in Figure 1(a) and supplementary \cite{Sup}), which is in close agreement with the literature \cite{Bruno2014,Bisogni2016}. A sharp and intense pre-peak at the O $K$ edge (528.8 eV) corresponds to the transition from O 1s core level to the ligand hole $\underline{\mathrm{L}}$ in the ${3d}^{7+\delta} \underline{\mathrm{L}}^\delta$ configuration (Figure 1(b)). Upon doping, we registered the following changes in the XAS spectra:

(1) A clear shift of the spectral weight from the high energy to the low energy component in the Ni-$L_3$ XAS profile. The Ni-$L_{2,3}$ edge position also shifts to lower energy by about 0.5 eV, from the undoped sample to the highest doping level. In this high doping limit, the XAS spectra are reminiscent of NiO where Ni has a 2+ oxidation state, strongly suggesting that doped carriers have been injected into the Ni conduction band.

(2) The pre-peak at the O $K$ edge is progressively suppressed until it completely disappears upon doping, indicating that doped carriers reside on the Ni $3d$ orbitals as well as the O ligand band. The disappearance of the XAS pre-peak in the highest doping sample suggests the filling of the ligand band upon doping, which resembles the spectra of LaNiO$_{2.5}$ \cite{Abbate2002}.

(3) The NNO spectra manifest a similar trend as SNO. Further details are reported in the supplementary materials \cite{Sup}.

\begin{figure}
	\includegraphics[width = 1 \columnwidth]{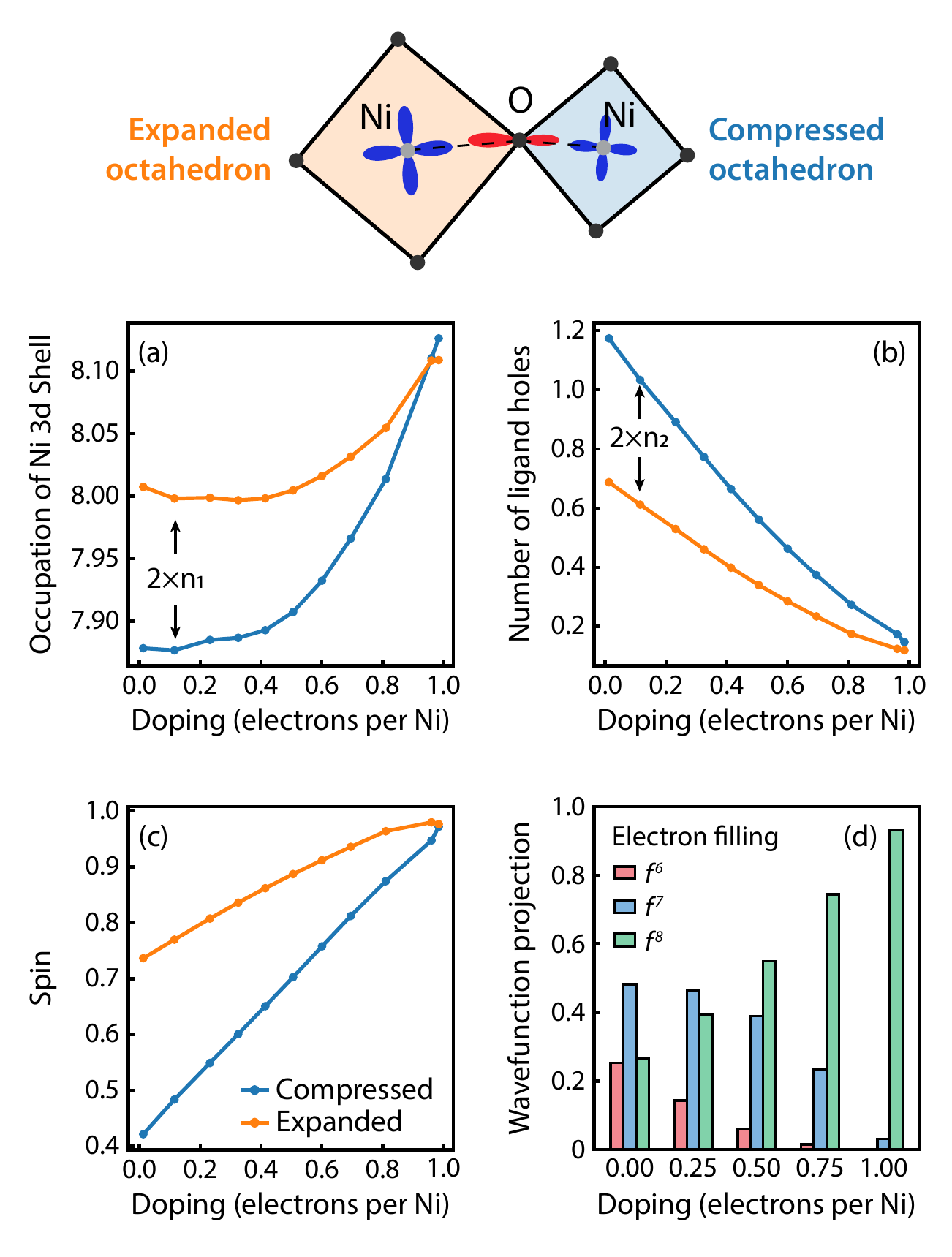}
	\caption{\label{fig2} (top) Double cluster electronic model for RENiO$_3$. The alternating expanded/compressed NiO$_6$ octahedra, inequivalent Ni $3d$ orbital occupation, and unbalanced O $2p$ orbitals are highlighted. (a-c) Doping evolution of the SmNiO$_{3-x}$ electronic structure from double cluster simulation. The evolution of different physical quantities for the two inequivalent sites are shown: (a) Occupation of Ni $3d$ orbitals; (b) Number of ligand holes; (c) Magnitude of Ni spin moments. The charge/bond disproportionation magnitudes as defined in equation (1) are marked out by arrows in (a,b). (d) The configuration weights of the ground-state wave function decomposed into states with different total electron filling ($f^6$, $f^7$, $f^8$) as a function of doping.}
\end{figure} 

To elucidate how the doped carriers are distributed in this correlated electronic ground state, we developed an extended multiplet ligand field theory, capable of modeling the ground state properties as well as the XAS of the doped system. Expanding on a previously successful quantum many body double cluster model \cite{Green2016}, we have added a charge reservoir term in the Hamiltonian which can be used to control the electron filling in the model (see supplemental materials for a complete description of the model \cite{Sup}). The calculations were implemented using the software QUANTY \cite{Haverkort2012,Haverkort2016}. 
The simulated SNO XAS spectra are overlaid onto the experimental data in Fig. 1(a)
for different doping levels (labeled according to the number of doped electrons per Ni atom). One can see that the simulated spectra capture all of the features and doping trends measured by XAS. By means of a least-squares best-fit analysis of the Ni-$L_3$ XAS edge experimental spectra vs. simulated ones, we can infer the doping levels for each sample shown in Figure 1(a). We note that due to self absorption effects, which artificially modulate the relative fluorescence yield at the $L_3$ and $L_2$ edges, a $\sim 30$\% discrepancy is found between the simulated and measured data at the Ni-$L_2$ edge \cite{Eisebitt1993}.

Figure 2 summarizes the effect of electron doping on the SNO electronic structure as captured by the doped double cluster simulation. Upon doping electrons into the system, we expect the extra carriers to redistribute in the oxygen ligand band and Ni $3d$ levels. Figure 2(a,b) shows the occupation of Ni $3d$ orbitals for the two inequivalent Ni sites vs. doping. The doped charges mostly occupy the oxygen ligand orbitals, whereas the Ni $3d$ orbitals begin filling only when doping exceeds $\sim$ 0.5 e$^{-}$/Ni. The difference in the occupation number between the two sites corresponds to the magnitude of the charge disproportionation ($n_1$). We note that there is a small amount of charge disproportionation in the undoped SNO sample. The charge disproportionation is found to be initially stable but strongly reduced when doping exceeds 0.5 e$^{-}$/Ni. In contrast, the strong bond disproportionation ($n_2$) presented in the different ligand hole occupation is continuously suppressed to zero upon doping as shown in Fig. 2(b). The doping also gradually changes the spin moments at both Ni sites from low to high spin states [Fig. 2(c)], consistent with previous evidence \cite{Wang2016}.

The doping-induced carrier redistribution and drastic changes to covalency were investigated by decomposing the ground-state many-electron wave function $|\psi\rangle = \Sigma_{n,i} ~ c_{n,i} |d^{n+i}\underline{L}^i\rangle$ into different Hilbert sub-spaces $f^n$ (spanned by basis vector $|d^{n}\underline{L}^0\rangle$, $|d^{n+1}\underline{L}^1\rangle$, $|d^{n+2}\underline{L}^2\rangle$...). The doping evolution of the configuration weight $\Sigma_{i} ~ c^2_{n,i} $ averaged between compressed and expanded octahedra for sub space $f^n$ where n = 6, 7, 8 is shown in Fig. 2(d). The undoped ground-state wave function has significant components in all three sub-spaces, indicating a highly covalent state. Upon doping, the system drastically loses covalency, and the ground-state wave function is dominated by single $f^8$ configuration. More details about the decomposition of the ground-state wave function into different basis can be found in supplementary material \cite{Sup}.

\begin{figure}
	\includegraphics[width = 1 \columnwidth]{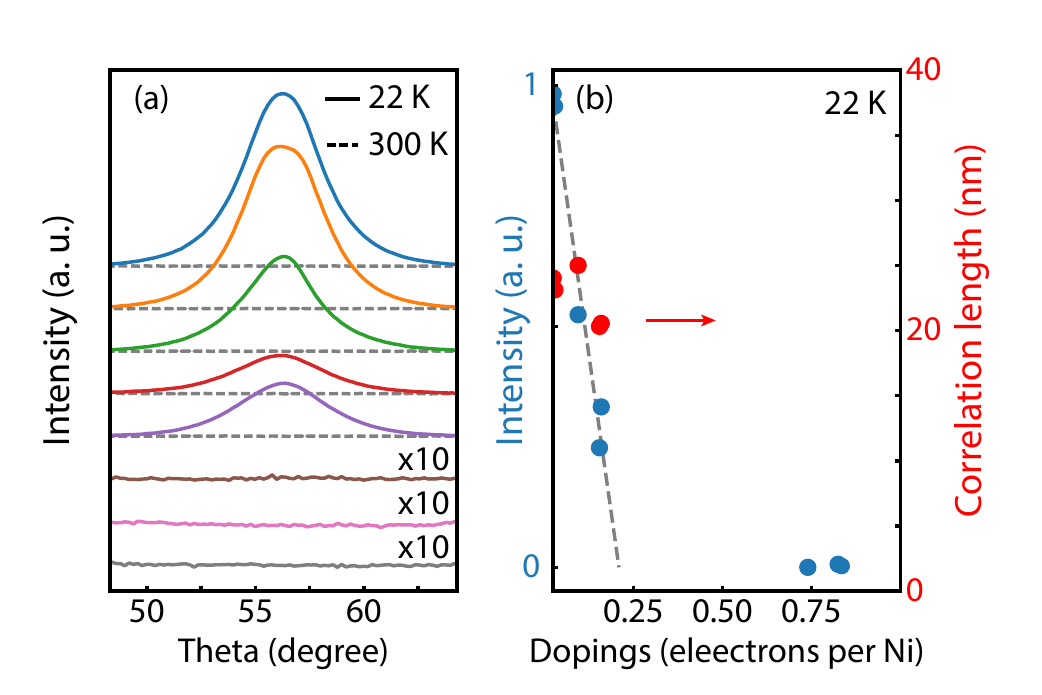}
	\caption{\label{fig3} (a) Rocking curves as a function of sample angle theta, across the $\mathbf{Q}_{AFM} = (1/4, 1/4, 1/4)_{pc}$ AFM reflection in SmNiO$_{3-x}$ at 22 K (colored solid line) and 300 K (grey dash line). The same color legend is used as the XAS data series in Figure 1. Curves are vertically offset for clarity. The intensity of the last three curves was rescaled to highlight that no scattering signal could be detected above the noise level. (b) Doping dependence of the magnetic Bragg peak intensity and correlation length measured at 22\,K. The dashed line is a linear fit to the data points from samples with AFM order and extrapolates to a threshold doping $n = 0.21$ for the sudden collapse of magnetic order.}
\end{figure}

We then turned our attention to the (1/4, 1/4, 1/4)$_{\mathrm{pc}}$ AFM order and its doping dependence. Fig. 3(a) shows the rocking curve across the (1/4, 1/4, 1/4)$_{\mathrm{pc}}$ magnetic superlattice peak below (solid line) and above (dashed line) the transition temperature for different doping levels and with the incident photon energy tuned at the Ni-$L_{3}$ resonance (853.2 eV). In the low doping region ($n<$0.3), a diffuse magnetic peak is found in all samples at 22 K, gradually decreasing at higher temperatures until it disappears upon warming above the transition temperature. The integrated AFM superlattice peak intensity decreases linearly in the low doping region [Fig. 3(b)], in contrast to the increase of Ni spin moments obtained from the simulation [Fig. 2(c)]. Despite the suppression of the total scattering intensity, no significant changes in the peak width or shape are observed, suggesting that the suppression of AFM order is not due to the creation of topological defects. The introduction of oxygen vacancies does not alter the thermodynamic properties of the AFM order, as the normalized temperature dependence of the integrated peak intensity are highly overlapped for the first five doping levels [Figure 4(a)]. No magnetic scattering intensity was observed above the noise level for higher doping levels (n$>$0.6) and down to 22 K, the lowest temperature measured in the present study. We note that the suppression of the (1/4, 1/4, 1/4)$_{\mathrm{pc}}$ AFM order does not preclude the emergence of magnetic order with different ordering vectors as previously found in the oxygen reduced nickelates \cite{Moriga1994, Sanchez1996, Alonso1997}.

The temperature-doping phase diagram is sketched out in Figure 4(b). The (1/4, 1/4, 1/4)$_{\mathrm{pc}}$ magnetic order is suppressed upon doping and collapses beyond a doping level of $\sim$ 0.21, as determined by a linear extrapolation of the intensity-doping relation for the low-doping samples, while the AFM transition temperature remains almost unchanged.

Oxygen vacancies are known to suppress the magnetic ordering temperature in manganites and cobaltites \cite{Trukhanov2002, Zhong2003, Balamurugan2006}, or give rise to new magnetic phases through ordering of oxygen defects \cite{Ramezanipour2012,Wang2018}. However, a reduction of the magnetic order parameter with no significant variation in the ordering temperature is unreported. Several scenarios are examined to explain the simultaneous increase of the Ni spin moment and decrease of the AFM order parameter, while $T_{AFM}$ remains unchanged. First, a microscopic phase separation picture may be invoked to explain the experimental results: the inhomogeneous distribution of oxygen vacancies creates two phases, with undoped AFM regions coexisting alongside doped non-magnetic ones. Upon doping, the AFM scattering intensity decreases linearly as the coverage of undoped regions is reduced, while the $T_{AFM}$ remains unchanged. To assess this possibility, we have performed a spectromicroscopy study using X-ray photoemission electron microscopy (XPEEM). No systematic electronic inhomogeneity was observed at either the O-$K$ edge or Ni-$L$ edge, down to the length scale of our spatial resolution limit ($\sim$ 10 nm), indicating an homogeneous electronic state with spatially uniform carrier doping. Details of the XPEEM data are described in the supplementary materials \cite{Sup}.

\begin{figure}
	\includegraphics[width = 1 \columnwidth]{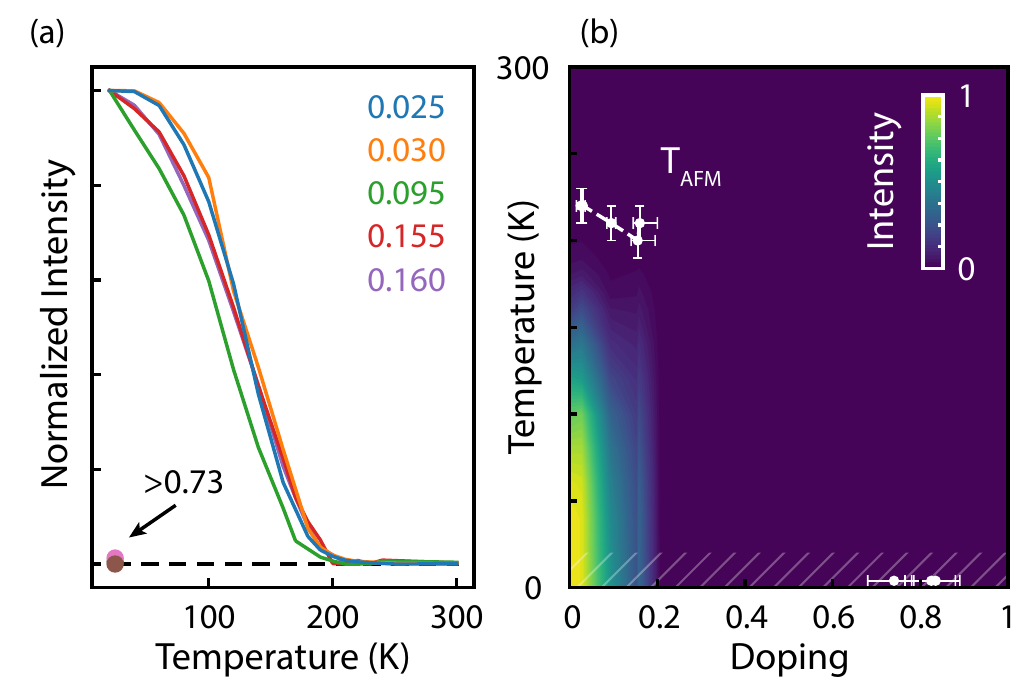}
	\caption{\label{fig4} (a) Temperature dependence of the AFM ordering peak integrated intensity in SmNiO$_{3-x}$. The intensity for the magnetic ordered samples are normalized to the lowest temperature. The same color legend is used as the XAS data series in Figure 1. (b) Temperature-doping plot of the magnetic phase diagram in SmNiO$_{3-x}$. The magnetic scattering intensity is color coded in the background. The AFM transition temperature for each sample is marked out. 
	The striped area highlights the temperature range that cannot be accessed in our study.
	}
\end{figure}

With a phase segregation scenario ruled out, we focused on an atomistic picture to explain the phase diagram. In this picture, the superexchange interaction between neighboring Ni spins is mediated by oxygen ligands. Each Ni atom is linked to its six nearest-neighbor Ni sites via six octahedrally coordinated oxygen atoms. The removal of oxygen not only alters the local Ni charges and spin moments via an effective doping mechanism, but it also destroys the superexchange interaction pathways that mediate the magnetic interaction across Ni moments. When the density of oxygen vacancies is low, long-range AFM order can still be sustained. When the atomic-scale disruption of the 3D magnetic superexchange network reaches a given threshold, long range magnetic order can no longer be supported. In our study, AFM order disappears at a doping level of around 0.21 electrons/Ni, corresponding to SmNiO$_{2.90}$.

In summary, we have systematically studied the electronic and magnetic structure of RENiO$_{3-x}$ (Re = Sm, Nd). The introduction of oxygen vacancies is shown to be an effective approach to continuously tune the ${3d}^{7+\delta}$\underline{L}$^\delta$ electronic ground state. By a combination of X-ray spectroscopy and model simulations, we established a protocol to quantitatively determine the doping level of each sample. The doped electrons are shown to redistribute into the Ni $3d$ level and O ligand hole bands and quickly suppress the charge disproportionation while leaving the bond disproportionation and local moments remarkably stable. We also show that electron doping has only marginal effect on (1/4, 1/4, 1/4)$_{\mathrm{pc}}$ AFM order except for a suppression of the ordering strength. The magnetic order collapses around a doping threshold of $n \sim 0.21$. No evidence of nanoscale phase separation is observed. We explain the evolution of magnetism as resulting from the progressive disruption of the superexchange interaction network upon introduction of oxygen vacancies. The results underscore the twofold role of oxygen vacancies in RENiO$_{3-x}$. On the one hand, the oxygen vacancies act as electron donors and homogeneously doped electrons into the valence band orbital manifold. On the other hand, the oxygen vacancies suppress magnetic order as they create local defects in the magnetic superexchange fabric. Our study reveals a surprising insensitivity of AFM ordering upon carrier doping, that singles out doped rare earth nickelates from the established phenomenology of other TMOs. Further, the sharp erasure of magnetic order realizes the possibility of reversible magneto-ionic switching of magnetism and use of oxygen-deficient nickelates for antiferromagnetic spintronics, low-power logic devices, and nonvolatile memory cells \cite{Jungwirth2016}.

We wish to thank George A. Sawatzky, Bernhard Keimer, Eva Benckiser, Matthias Hepting, Alex Fra\~{n}o, Alex McLeod, William Zheng, and John Mitchell for insightful discussions. 
This material is based upon work supported by the National Science Foundation under Grant No. 1751739. This work was supported by the Air Force Office of Scientific Research Young Investigator Program under grant FA9550-19-1-0063. RJG was supported by the Natural Sciences and Engineering Research Council of Canada (NSERC). Part of the research described in this paper was performed at the Canadian Light Source, a national research facility of the University of Saskatchewan, which is supported by the Canada Foundation for Innovation (CFI), NSERC, the National Research Council (NRC), the Canadian Institutes of Health Research (CIHR), the Government of Saskatchewan, and the University of Saskatchewan. This research used resources of the Center for Functional Nanomaterials and National Synchrotron Light Source II, which are US Department of Energy Office of Science Facilities at Brookhaven National Laboratory under Contract DE-SC0012704. S.R. acknowledges AFOSR grant FA9550-19-1-0351 for support.

\bibliography{OSNO}

\begin{thebibliography}{48}%
\makeatletter
\providecommand \@ifxundefined [1]{%
 \@ifx{#1\undefined}
}%
\providecommand \@ifnum [1]{%
 \ifnum #1\expandafter \@firstoftwo
 \else \expandafter \@secondoftwo
 \fi
}%
\providecommand \@ifx [1]{%
 \ifx #1\expandafter \@firstoftwo
 \else \expandafter \@secondoftwo
 \fi
}%
\providecommand \natexlab [1]{#1}%
\providecommand \enquote  [1]{``#1''}%
\providecommand \bibnamefont  [1]{#1}%
\providecommand \bibfnamefont [1]{#1}%
\providecommand \citenamefont [1]{#1}%
\providecommand \href@noop [0]{\@secondoftwo}%
\providecommand \href [0]{\begingroup \@sanitize@url \@href}%
\providecommand \@href[1]{\@@startlink{#1}\@@href}%
\providecommand \@@href[1]{\endgroup#1\@@endlink}%
\providecommand \@sanitize@url [0]{\catcode `\\12\catcode `\$12\catcode
  `\&12\catcode `\#12\catcode `\^12\catcode `\_12\catcode `\%12\relax}%
\providecommand \@@startlink[1]{}%
\providecommand \@@endlink[0]{}%
\providecommand \url  [0]{\begingroup\@sanitize@url \@url }%
\providecommand \@url [1]{\endgroup\@href {#1}{\urlprefix }}%
\providecommand \urlprefix  [0]{URL }%
\providecommand \Eprint [0]{\href }%
\providecommand \doibase [0]{http://dx.doi.org/}%
\providecommand \selectlanguage [0]{\@gobble}%
\providecommand \bibinfo  [0]{\@secondoftwo}%
\providecommand \bibfield  [0]{\@secondoftwo}%
\providecommand \translation [1]{[#1]}%
\providecommand \BibitemOpen [0]{}%
\providecommand \bibitemStop [0]{}%
\providecommand \bibitemNoStop [0]{.\EOS\space}%
\providecommand \EOS [0]{\spacefactor3000\relax}%
\providecommand \BibitemShut  [1]{\csname bibitem#1\endcsname}%
\let\auto@bib@innerbib\@empty
\bibitem [{\citenamefont {Fradkin}\ \emph {et~al.}(2015)\citenamefont
  {Fradkin}, \citenamefont {Kivelson},\ and\ \citenamefont
  {Tranquada}}]{Fradkin2015}%
  \BibitemOpen
  \bibfield  {author} {\bibinfo {author} {\bibfnamefont {E.}~\bibnamefont
  {Fradkin}}, \bibinfo {author} {\bibfnamefont {S.~A.}\ \bibnamefont
  {Kivelson}}, \ and\ \bibinfo {author} {\bibfnamefont {J.~M.}\ \bibnamefont
  {Tranquada}},\ }\href {\doibase 10.1103/RevModPhys.87.457} {\bibfield
  {journal} {\bibinfo  {journal} {Rev. Mod. Phys.}\ }\textbf {\bibinfo {volume}
  {87}},\ \bibinfo {pages} {457} (\bibinfo {year} {2015})}\BibitemShut
  {NoStop}%
\bibitem [{\citenamefont {Dagotto}(2005)}]{Dagotto2005}%
  \BibitemOpen
  \bibfield  {author} {\bibinfo {author} {\bibfnamefont {E.}~\bibnamefont
  {Dagotto}},\ }\href {\doibase 10.1126/science.1107559} {\bibfield  {journal}
  {\bibinfo  {journal} {Science}\ }\textbf {\bibinfo {volume} {309}},\ \bibinfo
  {pages} {257} (\bibinfo {year} {2005})}\BibitemShut {NoStop}%
\bibitem [{\citenamefont {Keimer}\ \emph {et~al.}(2015)\citenamefont {Keimer},
  \citenamefont {Kivelson}, \citenamefont {Norman}, \citenamefont {Uchida},\
  and\ \citenamefont {Zaanen}}]{Keimer2015}%
  \BibitemOpen
  \bibfield  {author} {\bibinfo {author} {\bibfnamefont {B.}~\bibnamefont
  {Keimer}}, \bibinfo {author} {\bibfnamefont {S.~A.}\ \bibnamefont
  {Kivelson}}, \bibinfo {author} {\bibfnamefont {M.~R.}\ \bibnamefont
  {Norman}}, \bibinfo {author} {\bibfnamefont {S.}~\bibnamefont {Uchida}}, \
  and\ \bibinfo {author} {\bibfnamefont {J.}~\bibnamefont {Zaanen}},\ }\href
  {\doibase 10.1038/nature14165} {\bibfield  {journal} {\bibinfo  {journal}
  {Nature}\ }\textbf {\bibinfo {volume} {518}},\ \bibinfo {pages} {179}
  (\bibinfo {year} {2015})}\BibitemShut {NoStop}%
\bibitem [{\citenamefont {Tallon}()}]{Tallon}%
  \BibitemOpen
  \bibfield  {author} {\bibinfo {author} {\bibfnamefont {J.~L.}\ \bibnamefont
  {Tallon}},\ }in\ \href {\doibase 10.1007/3-540-27294-1_7} {\emph {\bibinfo
  {booktitle} {Frontiers in Superconducting Materials}}}\ (\bibinfo
  {publisher} {Springer-Verlag},\ \bibinfo {address} {Berlin/Heidelberg})\ pp.\
  \bibinfo {pages} {295--330}\BibitemShut {NoStop}%
\bibitem [{\citenamefont {Herranz}\ \emph {et~al.}(2007)\citenamefont
  {Herranz}, \citenamefont {Basleti\ifmmode~\acute{c}\else \'{c}\fi{}},
  \citenamefont {Bibes}, \citenamefont {Carr\'et\'ero}, \citenamefont {Tafra},
  \citenamefont {Jacquet}, \citenamefont {Bouzehouane}, \citenamefont
  {Deranlot}, \citenamefont {Hamzi\ifmmode~\acute{c}\else \'{c}\fi{}},
  \citenamefont {Broto}, \citenamefont {Barth\'el\'emy},\ and\ \citenamefont
  {Fert}}]{Herranz2007}%
  \BibitemOpen
  \bibfield  {author} {\bibinfo {author} {\bibfnamefont {G.}~\bibnamefont
  {Herranz}}, \bibinfo {author} {\bibfnamefont {M.}~\bibnamefont
  {Basleti\ifmmode~\acute{c}\else \'{c}\fi{}}}, \bibinfo {author}
  {\bibfnamefont {M.}~\bibnamefont {Bibes}}, \bibinfo {author} {\bibfnamefont
  {C.}~\bibnamefont {Carr\'et\'ero}}, \bibinfo {author} {\bibfnamefont
  {E.}~\bibnamefont {Tafra}}, \bibinfo {author} {\bibfnamefont
  {E.}~\bibnamefont {Jacquet}}, \bibinfo {author} {\bibfnamefont
  {K.}~\bibnamefont {Bouzehouane}}, \bibinfo {author} {\bibfnamefont
  {C.}~\bibnamefont {Deranlot}}, \bibinfo {author} {\bibfnamefont
  {A.}~\bibnamefont {Hamzi\ifmmode~\acute{c}\else \'{c}\fi{}}}, \bibinfo
  {author} {\bibfnamefont {J.-M.}\ \bibnamefont {Broto}}, \bibinfo {author}
  {\bibfnamefont {A.}~\bibnamefont {Barth\'el\'emy}}, \ and\ \bibinfo {author}
  {\bibfnamefont {A.}~\bibnamefont {Fert}},\ }\href {\doibase
  10.1103/PhysRevLett.98.216803} {\bibfield  {journal} {\bibinfo  {journal}
  {Phys. Rev. Lett.}\ }\textbf {\bibinfo {volume} {98}},\ \bibinfo {pages}
  {216803} (\bibinfo {year} {2007})}\BibitemShut {NoStop}%
\bibitem [{\citenamefont {Zeng}\ \emph {et~al.}(1999)\citenamefont {Zeng},
  \citenamefont {Greenblatt},\ and\ \citenamefont {Croft}}]{Croft1999}%
  \BibitemOpen
  \bibfield  {author} {\bibinfo {author} {\bibfnamefont {Z.}~\bibnamefont
  {Zeng}}, \bibinfo {author} {\bibfnamefont {M.}~\bibnamefont {Greenblatt}}, \
  and\ \bibinfo {author} {\bibfnamefont {M.}~\bibnamefont {Croft}},\ }\href
  {\doibase 10.1103/PhysRevB.59.8784} {\bibfield  {journal} {\bibinfo
  {journal} {Phys. Rev. B}\ }\textbf {\bibinfo {volume} {59}},\ \bibinfo
  {pages} {8784} (\bibinfo {year} {1999})}\BibitemShut {NoStop}%
\bibitem [{\citenamefont {S\'anchez}\ \emph {et~al.}(1996)\citenamefont
  {S\'anchez}, \citenamefont {Causa}, \citenamefont {Caneiro}, \citenamefont
  {Butera}, \citenamefont {Vallet-Reg\'{\i}}, \citenamefont {Sayagu\'es},
  \citenamefont {Gonz\'alez-Calbet}, \citenamefont {Garc\'{\i}a-Sanz},\ and\
  \citenamefont {Rivas}}]{Sanchez1996}%
  \BibitemOpen
  \bibfield  {author} {\bibinfo {author} {\bibfnamefont {R.~D.}\ \bibnamefont
  {S\'anchez}}, \bibinfo {author} {\bibfnamefont {M.~T.}\ \bibnamefont
  {Causa}}, \bibinfo {author} {\bibfnamefont {A.}~\bibnamefont {Caneiro}},
  \bibinfo {author} {\bibfnamefont {A.}~\bibnamefont {Butera}}, \bibinfo
  {author} {\bibfnamefont {M.}~\bibnamefont {Vallet-Reg\'{\i}}}, \bibinfo
  {author} {\bibfnamefont {M.~J.}\ \bibnamefont {Sayagu\'es}}, \bibinfo
  {author} {\bibfnamefont {J.}~\bibnamefont {Gonz\'alez-Calbet}}, \bibinfo
  {author} {\bibfnamefont {F.}~\bibnamefont {Garc\'{\i}a-Sanz}}, \ and\
  \bibinfo {author} {\bibfnamefont {J.}~\bibnamefont {Rivas}},\ }\href
  {\doibase 10.1103/PhysRevB.54.16574} {\bibfield  {journal} {\bibinfo
  {journal} {Phys. Rev. B}\ }\textbf {\bibinfo {volume} {54}},\ \bibinfo
  {pages} {16574} (\bibinfo {year} {1996})}\BibitemShut {NoStop}%
\bibitem [{\citenamefont {Gouget}\ \emph {et~al.}(2019)\citenamefont {Gouget},
  \citenamefont {Duttine}, \citenamefont {Chung}, \citenamefont {Fourcade},
  \citenamefont {Mauvy}, \citenamefont {Braida}, \citenamefont {{Le Mercier}},\
  and\ \citenamefont {Demourgues}}]{Gouget2019}%
  \BibitemOpen
  \bibfield  {author} {\bibinfo {author} {\bibfnamefont {G.}~\bibnamefont
  {Gouget}}, \bibinfo {author} {\bibfnamefont {M.}~\bibnamefont {Duttine}},
  \bibinfo {author} {\bibfnamefont {U.~C.}\ \bibnamefont {Chung}}, \bibinfo
  {author} {\bibfnamefont {S.}~\bibnamefont {Fourcade}}, \bibinfo {author}
  {\bibfnamefont {F.}~\bibnamefont {Mauvy}}, \bibinfo {author} {\bibfnamefont
  {M.~D.}\ \bibnamefont {Braida}}, \bibinfo {author} {\bibfnamefont
  {T.}~\bibnamefont {{Le Mercier}}}, \ and\ \bibinfo {author} {\bibfnamefont
  {A.}~\bibnamefont {Demourgues}},\ }\href {\doibase
  10.1021/acs.chemmater.8b05292} {\bibfield  {journal} {\bibinfo  {journal}
  {Chemistry of Materials}\ }\textbf {\bibinfo {volume} {31}},\ \bibinfo
  {pages} {2828} (\bibinfo {year} {2019})}\BibitemShut {NoStop}%
\bibitem [{\citenamefont {Kan}\ \emph {et~al.}(2005)\citenamefont {Kan},
  \citenamefont {Terashima}, \citenamefont {Kanda}, \citenamefont {Masuno},
  \citenamefont {Tanaka}, \citenamefont {Chu}, \citenamefont {Kan},
  \citenamefont {Ishizumi}, \citenamefont {Kanemitsu}, \citenamefont
  {Shimakawa},\ and\ \citenamefont {Takano}}]{Kan2005}%
  \BibitemOpen
  \bibfield  {author} {\bibinfo {author} {\bibfnamefont {D.}~\bibnamefont
  {Kan}}, \bibinfo {author} {\bibfnamefont {T.}~\bibnamefont {Terashima}},
  \bibinfo {author} {\bibfnamefont {R.}~\bibnamefont {Kanda}}, \bibinfo
  {author} {\bibfnamefont {A.}~\bibnamefont {Masuno}}, \bibinfo {author}
  {\bibfnamefont {K.}~\bibnamefont {Tanaka}}, \bibinfo {author} {\bibfnamefont
  {S.}~\bibnamefont {Chu}}, \bibinfo {author} {\bibfnamefont {H.}~\bibnamefont
  {Kan}}, \bibinfo {author} {\bibfnamefont {A.}~\bibnamefont {Ishizumi}},
  \bibinfo {author} {\bibfnamefont {Y.}~\bibnamefont {Kanemitsu}}, \bibinfo
  {author} {\bibfnamefont {Y.}~\bibnamefont {Shimakawa}}, \ and\ \bibinfo
  {author} {\bibfnamefont {M.}~\bibnamefont {Takano}},\ }\href {\doibase
  10.1038/nmat1498} {\bibfield  {journal} {\bibinfo  {journal} {Nature
  Materials}\ }\textbf {\bibinfo {volume} {4}},\ \bibinfo {pages} {816}
  (\bibinfo {year} {2005})}\BibitemShut {NoStop}%
\bibitem [{\citenamefont {Mochizuki}\ \emph {et~al.}(2005)\citenamefont
  {Mochizuki}, \citenamefont {Fujishiro},\ and\ \citenamefont
  {Minami}}]{Mochizuki2005}%
  \BibitemOpen
  \bibfield  {author} {\bibinfo {author} {\bibfnamefont {S.}~\bibnamefont
  {Mochizuki}}, \bibinfo {author} {\bibfnamefont {F.}~\bibnamefont
  {Fujishiro}}, \ and\ \bibinfo {author} {\bibfnamefont {S.}~\bibnamefont
  {Minami}},\ }\href {\doibase 10.1088/0953-8984/17/6/011} {\bibfield
  {journal} {\bibinfo  {journal} {Journal of Physics Condensed Matter}\
  }\textbf {\bibinfo {volume} {17}},\ \bibinfo {pages} {923} (\bibinfo {year}
  {2005})}\BibitemShut {NoStop}%
\bibitem [{\citenamefont {Chen}\ \emph {et~al.}(2018)\citenamefont {Chen},
  \citenamefont {Zhang}, \citenamefont {Wang}, \citenamefont {Jalan},
  \citenamefont {Chen},\ and\ \citenamefont {Hou}}]{Chen2018}%
  \BibitemOpen
  \bibfield  {author} {\bibinfo {author} {\bibfnamefont {L.}~\bibnamefont
  {Chen}}, \bibinfo {author} {\bibfnamefont {Y.}~\bibnamefont {Zhang}},
  \bibinfo {author} {\bibfnamefont {X.}~\bibnamefont {Wang}}, \bibinfo {author}
  {\bibfnamefont {B.}~\bibnamefont {Jalan}}, \bibinfo {author} {\bibfnamefont
  {S.}~\bibnamefont {Chen}}, \ and\ \bibinfo {author} {\bibfnamefont
  {Y.}~\bibnamefont {Hou}},\ }\href {\doibase 10.1021/acs.jpcc.8b00653}
  {\bibfield  {journal} {\bibinfo  {journal} {The Journal of Physical Chemistry
  C}\ }\textbf {\bibinfo {volume} {122}},\ \bibinfo {pages} {11482} (\bibinfo
  {year} {2018})}\BibitemShut {NoStop}%
\bibitem [{\citenamefont {Li}\ \emph {et~al.}(2019)\citenamefont {Li},
  \citenamefont {Lee}, \citenamefont {Wang}, \citenamefont {Osada},
  \citenamefont {Crossley}, \citenamefont {Lee}, \citenamefont {Cui},
  \citenamefont {Hikita},\ and\ \citenamefont {Hwang}}]{Li2019}%
  \BibitemOpen
  \bibfield  {author} {\bibinfo {author} {\bibfnamefont {D.}~\bibnamefont
  {Li}}, \bibinfo {author} {\bibfnamefont {K.}~\bibnamefont {Lee}}, \bibinfo
  {author} {\bibfnamefont {B.~Y.}\ \bibnamefont {Wang}}, \bibinfo {author}
  {\bibfnamefont {M.}~\bibnamefont {Osada}}, \bibinfo {author} {\bibfnamefont
  {S.}~\bibnamefont {Crossley}}, \bibinfo {author} {\bibfnamefont {H.~R.}\
  \bibnamefont {Lee}}, \bibinfo {author} {\bibfnamefont {Y.}~\bibnamefont
  {Cui}}, \bibinfo {author} {\bibfnamefont {Y.}~\bibnamefont {Hikita}}, \ and\
  \bibinfo {author} {\bibfnamefont {H.~Y.}\ \bibnamefont {Hwang}},\ }\href
  {\doibase 10.1038/s41586-019-1496-5} {\bibfield  {journal} {\bibinfo
  {journal} {Nature}\ }\textbf {\bibinfo {volume} {572}},\ \bibinfo {pages}
  {624} (\bibinfo {year} {2019})}\BibitemShut {NoStop}%
\bibitem [{\citenamefont {Hoffmann}\ \emph {et~al.}(2015)\citenamefont
  {Hoffmann}, \citenamefont {Borisov}, \citenamefont {Ostanin}, \citenamefont
  {Mertig}, \citenamefont {Hergert},\ and\ \citenamefont
  {Ernst}}]{Hoffmann2015}%
  \BibitemOpen
  \bibfield  {author} {\bibinfo {author} {\bibfnamefont {M.}~\bibnamefont
  {Hoffmann}}, \bibinfo {author} {\bibfnamefont {V.~S.}\ \bibnamefont
  {Borisov}}, \bibinfo {author} {\bibfnamefont {S.}~\bibnamefont {Ostanin}},
  \bibinfo {author} {\bibfnamefont {I.}~\bibnamefont {Mertig}}, \bibinfo
  {author} {\bibfnamefont {W.}~\bibnamefont {Hergert}}, \ and\ \bibinfo
  {author} {\bibfnamefont {A.}~\bibnamefont {Ernst}},\ }\href {\doibase
  10.1103/PhysRevB.92.094427} {\bibfield  {journal} {\bibinfo  {journal} {Phys.
  Rev. B}\ }\textbf {\bibinfo {volume} {92}},\ \bibinfo {pages} {094427}
  (\bibinfo {year} {2015})}\BibitemShut {NoStop}%
\bibitem [{\citenamefont {Li}\ \emph {et~al.}(2013)\citenamefont {Li},
  \citenamefont {Zhao}, \citenamefont {Wang}, \citenamefont {Tang},
  \citenamefont {Zhu}, \citenamefont {Lee}, \citenamefont {Cao}, \citenamefont
  {Cai}, \citenamefont {Guo}, \citenamefont {Wang}, \citenamefont {Ling},
  \citenamefont {Pi}, \citenamefont {Jin}, \citenamefont {Zhang}, \citenamefont
  {Wang}, \citenamefont {Wang}, \citenamefont {Ju},\ and\ \citenamefont
  {Yang}}]{Li2013}%
  \BibitemOpen
  \bibfield  {author} {\bibinfo {author} {\bibfnamefont {W.}~\bibnamefont
  {Li}}, \bibinfo {author} {\bibfnamefont {R.}~\bibnamefont {Zhao}}, \bibinfo
  {author} {\bibfnamefont {L.}~\bibnamefont {Wang}}, \bibinfo {author}
  {\bibfnamefont {R.}~\bibnamefont {Tang}}, \bibinfo {author} {\bibfnamefont
  {Y.}~\bibnamefont {Zhu}}, \bibinfo {author} {\bibfnamefont {J.~H.}\
  \bibnamefont {Lee}}, \bibinfo {author} {\bibfnamefont {H.}~\bibnamefont
  {Cao}}, \bibinfo {author} {\bibfnamefont {T.}~\bibnamefont {Cai}}, \bibinfo
  {author} {\bibfnamefont {H.}~\bibnamefont {Guo}}, \bibinfo {author}
  {\bibfnamefont {C.}~\bibnamefont {Wang}}, \bibinfo {author} {\bibfnamefont
  {L.}~\bibnamefont {Ling}}, \bibinfo {author} {\bibfnamefont {L.}~\bibnamefont
  {Pi}}, \bibinfo {author} {\bibfnamefont {K.}~\bibnamefont {Jin}}, \bibinfo
  {author} {\bibfnamefont {Y.}~\bibnamefont {Zhang}}, \bibinfo {author}
  {\bibfnamefont {H.}~\bibnamefont {Wang}}, \bibinfo {author} {\bibfnamefont
  {Y.}~\bibnamefont {Wang}}, \bibinfo {author} {\bibfnamefont {S.}~\bibnamefont
  {Ju}}, \ and\ \bibinfo {author} {\bibfnamefont {H.}~\bibnamefont {Yang}},\
  }\href {\doibase 10.1038/srep02618} {\bibfield  {journal} {\bibinfo
  {journal} {Scientific Reports}\ }\textbf {\bibinfo {volume} {3}},\ \bibinfo
  {pages} {2618} (\bibinfo {year} {2013})}\BibitemShut {NoStop}%
\bibitem [{\citenamefont {Seikh}\ \emph {et~al.}(2008)\citenamefont {Seikh},
  \citenamefont {Simon}, \citenamefont {Caignaert}, \citenamefont {Pralong},
  \citenamefont {Lepetit}, \citenamefont {Boudin},\ and\ \citenamefont
  {Raveau}}]{Seikh2008}%
  \BibitemOpen
  \bibfield  {author} {\bibinfo {author} {\bibfnamefont {M.~M.}\ \bibnamefont
  {Seikh}}, \bibinfo {author} {\bibfnamefont {C.}~\bibnamefont {Simon}},
  \bibinfo {author} {\bibfnamefont {V.}~\bibnamefont {Caignaert}}, \bibinfo
  {author} {\bibfnamefont {V.}~\bibnamefont {Pralong}}, \bibinfo {author}
  {\bibfnamefont {M.~B.}\ \bibnamefont {Lepetit}}, \bibinfo {author}
  {\bibfnamefont {S.}~\bibnamefont {Boudin}}, \ and\ \bibinfo {author}
  {\bibfnamefont {B.}~\bibnamefont {Raveau}},\ }\href {\doibase
  10.1021/cm7026652} {\bibfield  {journal} {\bibinfo  {journal} {Chemistry of
  Materials}\ }\textbf {\bibinfo {volume} {20}},\ \bibinfo {pages} {231}
  (\bibinfo {year} {2008})}\BibitemShut {NoStop}%
\bibitem [{\citenamefont {Ramezanipour}\ \emph {et~al.}(2012)\citenamefont
  {Ramezanipour}, \citenamefont {Greedan}, \citenamefont {Siewenie},
  \citenamefont {Donaberger}, \citenamefont {Turner},\ and\ \citenamefont
  {Botton}}]{Ramezanipour2012}%
  \BibitemOpen
  \bibfield  {author} {\bibinfo {author} {\bibfnamefont {F.}~\bibnamefont
  {Ramezanipour}}, \bibinfo {author} {\bibfnamefont {J.~E.}\ \bibnamefont
  {Greedan}}, \bibinfo {author} {\bibfnamefont {J.}~\bibnamefont {Siewenie}},
  \bibinfo {author} {\bibfnamefont {R.~L.}\ \bibnamefont {Donaberger}},
  \bibinfo {author} {\bibfnamefont {S.}~\bibnamefont {Turner}}, \ and\ \bibinfo
  {author} {\bibfnamefont {G.~A.}\ \bibnamefont {Botton}},\ }\href {\doibase
  10.1021/ic202590r} {\bibfield  {journal} {\bibinfo  {journal} {Inorganic
  Chemistry}\ }\textbf {\bibinfo {volume} {51}},\ \bibinfo {pages} {2638}
  (\bibinfo {year} {2012})}\BibitemShut {NoStop}%
\bibitem [{\citenamefont {Balamurugan}\ \emph {et~al.}(2006)\citenamefont
  {Balamurugan}, \citenamefont {Yamaura}, \citenamefont {Karki}, \citenamefont
  {Young}, \citenamefont {Arai},\ and\ \citenamefont
  {Takayama-Muromachi}}]{Balamurugan2006}%
  \BibitemOpen
  \bibfield  {author} {\bibinfo {author} {\bibfnamefont {S.}~\bibnamefont
  {Balamurugan}}, \bibinfo {author} {\bibfnamefont {K.}~\bibnamefont
  {Yamaura}}, \bibinfo {author} {\bibfnamefont {A.~B.}\ \bibnamefont {Karki}},
  \bibinfo {author} {\bibfnamefont {D.~P.}\ \bibnamefont {Young}}, \bibinfo
  {author} {\bibfnamefont {M.}~\bibnamefont {Arai}}, \ and\ \bibinfo {author}
  {\bibfnamefont {E.}~\bibnamefont {Takayama-Muromachi}},\ }\href {\doibase
  10.1103/PhysRevB.74.172406} {\bibfield  {journal} {\bibinfo  {journal} {Phys.
  Rev. B}\ }\textbf {\bibinfo {volume} {74}},\ \bibinfo {pages} {172406}
  (\bibinfo {year} {2006})}\BibitemShut {NoStop}%
\bibitem [{\citenamefont {Basov}\ and\ \citenamefont
  {Chubukov}(2011)}]{Basov2011}%
  \BibitemOpen
  \bibfield  {author} {\bibinfo {author} {\bibfnamefont {D.~N.}\ \bibnamefont
  {Basov}}\ and\ \bibinfo {author} {\bibfnamefont {A.~V.}\ \bibnamefont
  {Chubukov}},\ }\href {\doibase 10.1038/nphys1975} {\bibfield  {journal}
  {\bibinfo  {journal} {Nature Physics}\ }\textbf {\bibinfo {volume} {7}},\
  \bibinfo {pages} {272} (\bibinfo {year} {2011})}\BibitemShut {NoStop}%
\bibitem [{\citenamefont {St{\o}len}\ \emph {et~al.}(2006)\citenamefont
  {St{\o}len}, \citenamefont {Bakken},\ and\ \citenamefont
  {Mohn}}]{Stolen2006}%
  \BibitemOpen
  \bibfield  {author} {\bibinfo {author} {\bibfnamefont {S.}~\bibnamefont
  {St{\o}len}}, \bibinfo {author} {\bibfnamefont {E.}~\bibnamefont {Bakken}}, \
  and\ \bibinfo {author} {\bibfnamefont {C.~E.}\ \bibnamefont {Mohn}},\ }\href
  {\doibase 10.1039/b512271f} {\bibfield  {journal} {\bibinfo  {journal}
  {Physical Chemistry Chemical Physics}\ }\textbf {\bibinfo {volume} {8}},\
  \bibinfo {pages} {429} (\bibinfo {year} {2006})}\BibitemShut {NoStop}%
\bibitem [{\citenamefont {Catalan}(2008)}]{Catalan2008}%
  \BibitemOpen
  \bibfield  {author} {\bibinfo {author} {\bibfnamefont {G.}~\bibnamefont
  {Catalan}},\ }\href {\doibase 10.1080/01411590801992463} {\bibfield
  {journal} {\bibinfo  {journal} {Phase Transitions}\ }\textbf {\bibinfo
  {volume} {81}},\ \bibinfo {pages} {729} (\bibinfo {year} {2008})}\BibitemShut
  {NoStop}%
\bibitem [{\citenamefont {Middey}\ \emph {et~al.}(2016)\citenamefont {Middey},
  \citenamefont {Chakhalian}, \citenamefont {Mahadevan}, \citenamefont
  {Freeland}, \citenamefont {Millis},\ and\ \citenamefont
  {Sarma}}]{Middey2016}%
  \BibitemOpen
  \bibfield  {author} {\bibinfo {author} {\bibfnamefont {S.}~\bibnamefont
  {Middey}}, \bibinfo {author} {\bibfnamefont {J.}~\bibnamefont {Chakhalian}},
  \bibinfo {author} {\bibfnamefont {P.}~\bibnamefont {Mahadevan}}, \bibinfo
  {author} {\bibfnamefont {J.}~\bibnamefont {Freeland}}, \bibinfo {author}
  {\bibfnamefont {A.}~\bibnamefont {Millis}}, \ and\ \bibinfo {author}
  {\bibfnamefont {D.}~\bibnamefont {Sarma}},\ }\href {\doibase
  10.1146/annurev-matsci-070115-032057} {\bibfield  {journal} {\bibinfo
  {journal} {Annual Review of Materials Research}\ }\textbf {\bibinfo {volume}
  {46}},\ \bibinfo {pages} {305} (\bibinfo {year} {2016})}\BibitemShut
  {NoStop}%
\bibitem [{\citenamefont {Catalano}\ \emph {et~al.}(2018)\citenamefont
  {Catalano}, \citenamefont {Gibert}, \citenamefont {Fowlie}, \citenamefont
  {{\'{I}}{\~{n}}iguez}, \citenamefont {Triscone},\ and\ \citenamefont
  {Kreisel}}]{Catalano2018}%
  \BibitemOpen
  \bibfield  {author} {\bibinfo {author} {\bibfnamefont {S.}~\bibnamefont
  {Catalano}}, \bibinfo {author} {\bibfnamefont {M.}~\bibnamefont {Gibert}},
  \bibinfo {author} {\bibfnamefont {J.}~\bibnamefont {Fowlie}}, \bibinfo
  {author} {\bibfnamefont {J.}~\bibnamefont {{\'{I}}{\~{n}}iguez}}, \bibinfo
  {author} {\bibfnamefont {J.-M.}\ \bibnamefont {Triscone}}, \ and\ \bibinfo
  {author} {\bibfnamefont {J.}~\bibnamefont {Kreisel}},\ }\href {\doibase
  10.1088/1361-6633/aaa37a} {\bibfield  {journal} {\bibinfo  {journal} {Reports
  on Progress in Physics}\ }\textbf {\bibinfo {volume} {81}},\ \bibinfo {pages}
  {046501} (\bibinfo {year} {2018})}\BibitemShut {NoStop}%
\bibitem [{\citenamefont {Torrance}\ \emph {et~al.}(1992)\citenamefont
  {Torrance}, \citenamefont {Lacorre}, \citenamefont {Nazzal}, \citenamefont
  {Ansaldo},\ and\ \citenamefont {Niedermayer}}]{Torrance1992}%
  \BibitemOpen
  \bibfield  {author} {\bibinfo {author} {\bibfnamefont {J.~B.}\ \bibnamefont
  {Torrance}}, \bibinfo {author} {\bibfnamefont {P.}~\bibnamefont {Lacorre}},
  \bibinfo {author} {\bibfnamefont {A.~I.}\ \bibnamefont {Nazzal}}, \bibinfo
  {author} {\bibfnamefont {E.~J.}\ \bibnamefont {Ansaldo}}, \ and\ \bibinfo
  {author} {\bibfnamefont {C.}~\bibnamefont {Niedermayer}},\ }\href {\doibase
  10.1103/PhysRevB.45.8209} {\bibfield  {journal} {\bibinfo  {journal} {Phys.
  Rev. B}\ }\textbf {\bibinfo {volume} {45}},\ \bibinfo {pages} {8209}
  (\bibinfo {year} {1992})}\BibitemShut {NoStop}%
\bibitem [{\citenamefont {Zaanen}\ \emph {et~al.}(1985)\citenamefont {Zaanen},
  \citenamefont {Sawatzky},\ and\ \citenamefont {Allen}}]{Zaanen1985}%
  \BibitemOpen
  \bibfield  {author} {\bibinfo {author} {\bibfnamefont {J.}~\bibnamefont
  {Zaanen}}, \bibinfo {author} {\bibfnamefont {G.~A.}\ \bibnamefont
  {Sawatzky}}, \ and\ \bibinfo {author} {\bibfnamefont {J.~W.}\ \bibnamefont
  {Allen}},\ }\href {\doibase 10.1103/PhysRevLett.55.418} {\bibfield  {journal}
  {\bibinfo  {journal} {Phys. Rev. Lett.}\ }\textbf {\bibinfo {volume} {55}},\
  \bibinfo {pages} {418} (\bibinfo {year} {1985})}\BibitemShut {NoStop}%
\bibitem [{\citenamefont {Mizokawa}\ \emph {et~al.}(2000)\citenamefont
  {Mizokawa}, \citenamefont {Khomskii},\ and\ \citenamefont
  {Sawatzky}}]{Mizokawa1999}%
  \BibitemOpen
  \bibfield  {author} {\bibinfo {author} {\bibfnamefont {T.}~\bibnamefont
  {Mizokawa}}, \bibinfo {author} {\bibfnamefont {D.~I.}\ \bibnamefont
  {Khomskii}}, \ and\ \bibinfo {author} {\bibfnamefont {G.~A.}\ \bibnamefont
  {Sawatzky}},\ }\href {\doibase 10.1103/PhysRevB.61.11263} {\bibfield
  {journal} {\bibinfo  {journal} {Phys. Rev. B}\ }\textbf {\bibinfo {volume}
  {61}},\ \bibinfo {pages} {11263} (\bibinfo {year} {2000})}\BibitemShut
  {NoStop}%
\bibitem [{\citenamefont {Bisogni}\ \emph {et~al.}(2016)\citenamefont
  {Bisogni}, \citenamefont {Catalano}, \citenamefont {Green}, \citenamefont
  {Gibert}, \citenamefont {Scherwitzl}, \citenamefont {Huang}, \citenamefont
  {Strocov}, \citenamefont {Zubko}, \citenamefont {Balandeh}, \citenamefont
  {Triscone}, \citenamefont {Sawatzky},\ and\ \citenamefont
  {Schmitt}}]{Bisogni2016}%
  \BibitemOpen
  \bibfield  {author} {\bibinfo {author} {\bibfnamefont {V.}~\bibnamefont
  {Bisogni}}, \bibinfo {author} {\bibfnamefont {S.}~\bibnamefont {Catalano}},
  \bibinfo {author} {\bibfnamefont {R.~J.}\ \bibnamefont {Green}}, \bibinfo
  {author} {\bibfnamefont {M.}~\bibnamefont {Gibert}}, \bibinfo {author}
  {\bibfnamefont {R.}~\bibnamefont {Scherwitzl}}, \bibinfo {author}
  {\bibfnamefont {Y.}~\bibnamefont {Huang}}, \bibinfo {author} {\bibfnamefont
  {V.~N.}\ \bibnamefont {Strocov}}, \bibinfo {author} {\bibfnamefont
  {P.}~\bibnamefont {Zubko}}, \bibinfo {author} {\bibfnamefont
  {S.}~\bibnamefont {Balandeh}}, \bibinfo {author} {\bibfnamefont {J.-M.}\
  \bibnamefont {Triscone}}, \bibinfo {author} {\bibfnamefont {G.}~\bibnamefont
  {Sawatzky}}, \ and\ \bibinfo {author} {\bibfnamefont {T.}~\bibnamefont
  {Schmitt}},\ }\href {\doibase 10.1038/ncomms13017} {\bibfield  {journal}
  {\bibinfo  {journal} {Nature Communications}\ }\textbf {\bibinfo {volume}
  {7}},\ \bibinfo {pages} {13017} (\bibinfo {year} {2016})}\BibitemShut
  {NoStop}%
\bibitem [{\citenamefont {Garc{\'{\i}}a-Mu{\~{n}}oz}\ \emph
  {et~al.}(1992)\citenamefont {Garc{\'{\i}}a-Mu{\~{n}}oz}, \citenamefont
  {Rodr{\'{\i}}guez-Carvajal},\ and\ \citenamefont
  {Lacorre}}]{Gaecia-Munoz1992}%
  \BibitemOpen
  \bibfield  {author} {\bibinfo {author} {\bibfnamefont {J.~L.}\ \bibnamefont
  {Garc{\'{\i}}a-Mu{\~{n}}oz}}, \bibinfo {author} {\bibfnamefont
  {J.}~\bibnamefont {Rodr{\'{\i}}guez-Carvajal}}, \ and\ \bibinfo {author}
  {\bibfnamefont {P.}~\bibnamefont {Lacorre}},\ }\href {\doibase
  10.1209/0295-5075/20/3/009} {\bibfield  {journal} {\bibinfo  {journal}
  {Europhysics Letters ({EPL})}\ }\textbf {\bibinfo {volume} {20}},\ \bibinfo
  {pages} {241} (\bibinfo {year} {1992})}\BibitemShut {NoStop}%
\bibitem [{\citenamefont {Scagnoli}\ \emph {et~al.}(2006)\citenamefont
  {Scagnoli}, \citenamefont {Staub}, \citenamefont {Mulders}, \citenamefont
  {Janousch}, \citenamefont {Meijer}, \citenamefont {Hammerl}, \citenamefont
  {Tonnerre},\ and\ \citenamefont {Stojic}}]{Scagnoli2006}%
  \BibitemOpen
  \bibfield  {author} {\bibinfo {author} {\bibfnamefont {V.}~\bibnamefont
  {Scagnoli}}, \bibinfo {author} {\bibfnamefont {U.}~\bibnamefont {Staub}},
  \bibinfo {author} {\bibfnamefont {A.~M.}\ \bibnamefont {Mulders}}, \bibinfo
  {author} {\bibfnamefont {M.}~\bibnamefont {Janousch}}, \bibinfo {author}
  {\bibfnamefont {G.~I.}\ \bibnamefont {Meijer}}, \bibinfo {author}
  {\bibfnamefont {G.}~\bibnamefont {Hammerl}}, \bibinfo {author} {\bibfnamefont
  {J.~M.}\ \bibnamefont {Tonnerre}}, \ and\ \bibinfo {author} {\bibfnamefont
  {N.}~\bibnamefont {Stojic}},\ }\href {\doibase 10.1103/PhysRevB.73.100409}
  {\bibfield  {journal} {\bibinfo  {journal} {Phys. Rev. B}\ }\textbf {\bibinfo
  {volume} {73}},\ \bibinfo {pages} {100409} (\bibinfo {year}
  {2006})}\BibitemShut {NoStop}%
\bibitem [{\citenamefont {Frano}\ \emph {et~al.}(2013)\citenamefont {Frano},
  \citenamefont {Schierle}, \citenamefont {Haverkort}, \citenamefont {Lu},
  \citenamefont {Wu}, \citenamefont {Blanco-Canosa}, \citenamefont {Nwankwo},
  \citenamefont {Boris}, \citenamefont {Wochner}, \citenamefont {Cristiani},
  \citenamefont {Habermeier}, \citenamefont {Logvenov}, \citenamefont {Hinkov},
  \citenamefont {Benckiser}, \citenamefont {Weschke},\ and\ \citenamefont
  {Keimer}}]{Frano2013}%
  \BibitemOpen
  \bibfield  {author} {\bibinfo {author} {\bibfnamefont {A.}~\bibnamefont
  {Frano}}, \bibinfo {author} {\bibfnamefont {E.}~\bibnamefont {Schierle}},
  \bibinfo {author} {\bibfnamefont {M.~W.}\ \bibnamefont {Haverkort}}, \bibinfo
  {author} {\bibfnamefont {Y.}~\bibnamefont {Lu}}, \bibinfo {author}
  {\bibfnamefont {M.}~\bibnamefont {Wu}}, \bibinfo {author} {\bibfnamefont
  {S.}~\bibnamefont {Blanco-Canosa}}, \bibinfo {author} {\bibfnamefont
  {U.}~\bibnamefont {Nwankwo}}, \bibinfo {author} {\bibfnamefont {A.~V.}\
  \bibnamefont {Boris}}, \bibinfo {author} {\bibfnamefont {P.}~\bibnamefont
  {Wochner}}, \bibinfo {author} {\bibfnamefont {G.}~\bibnamefont {Cristiani}},
  \bibinfo {author} {\bibfnamefont {H.~U.}\ \bibnamefont {Habermeier}},
  \bibinfo {author} {\bibfnamefont {G.}~\bibnamefont {Logvenov}}, \bibinfo
  {author} {\bibfnamefont {V.}~\bibnamefont {Hinkov}}, \bibinfo {author}
  {\bibfnamefont {E.}~\bibnamefont {Benckiser}}, \bibinfo {author}
  {\bibfnamefont {E.}~\bibnamefont {Weschke}}, \ and\ \bibinfo {author}
  {\bibfnamefont {B.}~\bibnamefont {Keimer}},\ }\href {\doibase
  10.1103/PhysRevLett.111.106804} {\bibfield  {journal} {\bibinfo  {journal}
  {Phys. Rev. Lett.}\ }\textbf {\bibinfo {volume} {111}},\ \bibinfo {pages}
  {106804} (\bibinfo {year} {2013})}\BibitemShut {NoStop}%
\bibitem [{\citenamefont {Hepting}\ \emph {et~al.}(2018)\citenamefont
  {Hepting}, \citenamefont {Green}, \citenamefont {Zhong}, \citenamefont
  {Bluschke}, \citenamefont {Suyolcu}, \citenamefont {Macke}, \citenamefont
  {Frano}, \citenamefont {Catalano}, \citenamefont {Gibert}, \citenamefont
  {Sutarto}, \citenamefont {He}, \citenamefont {Cristiani}, \citenamefont
  {Logvenov}, \citenamefont {Wang}, \citenamefont {van Aken}, \citenamefont
  {Hansmann}, \citenamefont {Le~Tacon}, \citenamefont {Triscone}, \citenamefont
  {Sawatzky}, \citenamefont {Keimer},\ and\ \citenamefont
  {Benckiser}}]{Hepting2018}%
  \BibitemOpen
  \bibfield  {author} {\bibinfo {author} {\bibfnamefont {M.}~\bibnamefont
  {Hepting}}, \bibinfo {author} {\bibfnamefont {R.~J.}\ \bibnamefont {Green}},
  \bibinfo {author} {\bibfnamefont {Z.}~\bibnamefont {Zhong}}, \bibinfo
  {author} {\bibfnamefont {M.}~\bibnamefont {Bluschke}}, \bibinfo {author}
  {\bibfnamefont {Y.~E.}\ \bibnamefont {Suyolcu}}, \bibinfo {author}
  {\bibfnamefont {S.}~\bibnamefont {Macke}}, \bibinfo {author} {\bibfnamefont
  {A.}~\bibnamefont {Frano}}, \bibinfo {author} {\bibfnamefont
  {S.}~\bibnamefont {Catalano}}, \bibinfo {author} {\bibfnamefont
  {M.}~\bibnamefont {Gibert}}, \bibinfo {author} {\bibfnamefont
  {R.}~\bibnamefont {Sutarto}}, \bibinfo {author} {\bibfnamefont
  {F.}~\bibnamefont {He}}, \bibinfo {author} {\bibfnamefont {G.}~\bibnamefont
  {Cristiani}}, \bibinfo {author} {\bibfnamefont {G.}~\bibnamefont {Logvenov}},
  \bibinfo {author} {\bibfnamefont {Y.}~\bibnamefont {Wang}}, \bibinfo {author}
  {\bibfnamefont {P.~A.}\ \bibnamefont {van Aken}}, \bibinfo {author}
  {\bibfnamefont {P.}~\bibnamefont {Hansmann}}, \bibinfo {author}
  {\bibfnamefont {M.}~\bibnamefont {Le~Tacon}}, \bibinfo {author}
  {\bibfnamefont {J.-M.}\ \bibnamefont {Triscone}}, \bibinfo {author}
  {\bibfnamefont {G.~A.}\ \bibnamefont {Sawatzky}}, \bibinfo {author}
  {\bibfnamefont {B.}~\bibnamefont {Keimer}}, \ and\ \bibinfo {author}
  {\bibfnamefont {E.}~\bibnamefont {Benckiser}},\ }\href {\doibase
  10.1038/s41567-018-0218-5} {\bibfield  {journal} {\bibinfo  {journal} {Nature
  Physics}\ }\textbf {\bibinfo {volume} {14}},\ \bibinfo {pages} {1097}
  (\bibinfo {year} {2018})}\BibitemShut {NoStop}%
\bibitem [{\citenamefont {Wang}\ \emph {et~al.}(2016)\citenamefont {Wang},
  \citenamefont {Dash}, \citenamefont {Chang}, \citenamefont {You},
  \citenamefont {Feng}, \citenamefont {He}, \citenamefont {Jin}, \citenamefont
  {Zhou}, \citenamefont {Ong}, \citenamefont {Ren}, \citenamefont {Wang},
  \citenamefont {Chen},\ and\ \citenamefont {Wang}}]{Wang2016}%
  \BibitemOpen
  \bibfield  {author} {\bibinfo {author} {\bibfnamefont {L.}~\bibnamefont
  {Wang}}, \bibinfo {author} {\bibfnamefont {S.}~\bibnamefont {Dash}}, \bibinfo
  {author} {\bibfnamefont {L.}~\bibnamefont {Chang}}, \bibinfo {author}
  {\bibfnamefont {L.}~\bibnamefont {You}}, \bibinfo {author} {\bibfnamefont
  {Y.}~\bibnamefont {Feng}}, \bibinfo {author} {\bibfnamefont {X.}~\bibnamefont
  {He}}, \bibinfo {author} {\bibfnamefont {K.~J.}\ \bibnamefont {Jin}},
  \bibinfo {author} {\bibfnamefont {Y.}~\bibnamefont {Zhou}}, \bibinfo {author}
  {\bibfnamefont {H.~G.}\ \bibnamefont {Ong}}, \bibinfo {author} {\bibfnamefont
  {P.}~\bibnamefont {Ren}}, \bibinfo {author} {\bibfnamefont {S.}~\bibnamefont
  {Wang}}, \bibinfo {author} {\bibfnamefont {L.}~\bibnamefont {Chen}}, \ and\
  \bibinfo {author} {\bibfnamefont {J.}~\bibnamefont {Wang}},\ }\href {\doibase
  10.1021/acsami.6b00650} {\bibfield  {journal} {\bibinfo  {journal} {ACS
  Applied Materials and Interfaces}\ }\textbf {\bibinfo {volume} {8}},\
  \bibinfo {pages} {9769} (\bibinfo {year} {2016})}\BibitemShut {NoStop}%
\bibitem [{\citenamefont {Kotiuga}\ \emph {et~al.}(2019)\citenamefont
  {Kotiuga}, \citenamefont {Zhang}, \citenamefont {Li}, \citenamefont
  {Rodolakis}, \citenamefont {Zhou}, \citenamefont {Sutarto}, \citenamefont
  {He}, \citenamefont {Wang}, \citenamefont {Sun}, \citenamefont {Wang},
  \citenamefont {Aghamiri}, \citenamefont {Hancock}, \citenamefont {Rokhinson},
  \citenamefont {Landau}, \citenamefont {Abate}, \citenamefont {Freeland},
  \citenamefont {Comin}, \citenamefont {Ramanathan},\ and\ \citenamefont
  {Rabe}}]{Kotiuga2019}%
  \BibitemOpen
  \bibfield  {author} {\bibinfo {author} {\bibfnamefont {M.}~\bibnamefont
  {Kotiuga}}, \bibinfo {author} {\bibfnamefont {Z.}~\bibnamefont {Zhang}},
  \bibinfo {author} {\bibfnamefont {J.}~\bibnamefont {Li}}, \bibinfo {author}
  {\bibfnamefont {F.}~\bibnamefont {Rodolakis}}, \bibinfo {author}
  {\bibfnamefont {H.}~\bibnamefont {Zhou}}, \bibinfo {author} {\bibfnamefont
  {R.}~\bibnamefont {Sutarto}}, \bibinfo {author} {\bibfnamefont
  {F.}~\bibnamefont {He}}, \bibinfo {author} {\bibfnamefont {Q.}~\bibnamefont
  {Wang}}, \bibinfo {author} {\bibfnamefont {Y.}~\bibnamefont {Sun}}, \bibinfo
  {author} {\bibfnamefont {Y.}~\bibnamefont {Wang}}, \bibinfo {author}
  {\bibfnamefont {N.~A.}\ \bibnamefont {Aghamiri}}, \bibinfo {author}
  {\bibfnamefont {S.~B.}\ \bibnamefont {Hancock}}, \bibinfo {author}
  {\bibfnamefont {L.~P.}\ \bibnamefont {Rokhinson}}, \bibinfo {author}
  {\bibfnamefont {D.~P.}\ \bibnamefont {Landau}}, \bibinfo {author}
  {\bibfnamefont {Y.}~\bibnamefont {Abate}}, \bibinfo {author} {\bibfnamefont
  {J.~W.}\ \bibnamefont {Freeland}}, \bibinfo {author} {\bibfnamefont
  {R.}~\bibnamefont {Comin}}, \bibinfo {author} {\bibfnamefont
  {S.}~\bibnamefont {Ramanathan}}, \ and\ \bibinfo {author} {\bibfnamefont
  {K.~M.}\ \bibnamefont {Rabe}},\ }\href {\doibase 10.1073/pnas.1910490116}
  {\bibfield  {journal} {\bibinfo  {journal} {Proceedings of the National
  Academy of Sciences}\ }\textbf {\bibinfo {volume} {116}},\ \bibinfo {pages}
  {21992} (\bibinfo {year} {2019})}\BibitemShut {NoStop}%
\bibitem [{\citenamefont {Nikulin}\ \emph {et~al.}(2004)\citenamefont
  {Nikulin}, \citenamefont {Novojilov}, \citenamefont {Kaul}, \citenamefont
  {Mudretsova},\ and\ \citenamefont {Kondrashov}}]{Nikulin2004}%
  \BibitemOpen
  \bibfield  {author} {\bibinfo {author} {\bibfnamefont {I.~V.}\ \bibnamefont
  {Nikulin}}, \bibinfo {author} {\bibfnamefont {M.~A.}\ \bibnamefont
  {Novojilov}}, \bibinfo {author} {\bibfnamefont {A.~R.}\ \bibnamefont {Kaul}},
  \bibinfo {author} {\bibfnamefont {S.~N.}\ \bibnamefont {Mudretsova}}, \ and\
  \bibinfo {author} {\bibfnamefont {S.~V.}\ \bibnamefont {Kondrashov}},\ }\href
  {\doibase 10.1016/j.materresbull.2004.02.005} {\bibfield  {journal} {\bibinfo
   {journal} {Materials Research Bulletin}\ }\textbf {\bibinfo {volume} {39}},\
  \bibinfo {pages} {775} (\bibinfo {year} {2004})}\BibitemShut {NoStop}%
\bibitem [{\citenamefont {Tiwari}\ and\ \citenamefont
  {Rajeev}(1998)}]{Tiwari1999}%
  \BibitemOpen
  \bibfield  {author} {\bibinfo {author} {\bibfnamefont {A.}~\bibnamefont
  {Tiwari}}\ and\ \bibinfo {author} {\bibfnamefont {K.}~\bibnamefont
  {Rajeev}},\ }\href {\doibase 10.1016/S0038-1098(98)00515-8} {\bibfield
  {journal} {\bibinfo  {journal} {Solid State Communications}\ }\textbf
  {\bibinfo {volume} {109}},\ \bibinfo {pages} {119} (\bibinfo {year}
  {1998})}\BibitemShut {NoStop}%
\bibitem [{\citenamefont {Garc\'{\i}a-Mu\~noz}\ \emph
  {et~al.}(1995)\citenamefont {Garc\'{\i}a-Mu\~noz}, \citenamefont {Suaaidi},
  \citenamefont {Mart\'{\i}nez-Lope},\ and\ \citenamefont
  {Alonso}}]{Garcia-Munoz1995}%
  \BibitemOpen
  \bibfield  {author} {\bibinfo {author} {\bibfnamefont {J.~L.}\ \bibnamefont
  {Garc\'{\i}a-Mu\~noz}}, \bibinfo {author} {\bibfnamefont {M.}~\bibnamefont
  {Suaaidi}}, \bibinfo {author} {\bibfnamefont {M.~J.}\ \bibnamefont
  {Mart\'{\i}nez-Lope}}, \ and\ \bibinfo {author} {\bibfnamefont {J.~A.}\
  \bibnamefont {Alonso}},\ }\href {\doibase 10.1103/PhysRevB.52.13563}
  {\bibfield  {journal} {\bibinfo  {journal} {Phys. Rev. B}\ }\textbf {\bibinfo
  {volume} {52}},\ \bibinfo {pages} {13563} (\bibinfo {year}
  {1995})}\BibitemShut {NoStop}%
\bibitem [{Sup()}]{Sup}%
  \BibitemOpen
  \href@noop {} {\bibinfo  {journal} {See Supplemental Material for additional
  details}\ }\BibitemShut {NoStop}%
\bibitem [{\citenamefont {Bruno}\ \emph {et~al.}(2014)\citenamefont {Bruno},
  \citenamefont {Valencia}, \citenamefont {Abrudan}, \citenamefont {Dumont},
  \citenamefont {Carrétéro}, \citenamefont {Bibes},\ and\ \citenamefont
  {Barthélémy}}]{Bruno2014}%
  \BibitemOpen
\bibfield  {journal} {  }\bibfield  {author} {\bibinfo {author} {\bibfnamefont
  {F.~Y.}\ \bibnamefont {Bruno}}, \bibinfo {author} {\bibfnamefont
  {S.}~\bibnamefont {Valencia}}, \bibinfo {author} {\bibfnamefont
  {R.}~\bibnamefont {Abrudan}}, \bibinfo {author} {\bibfnamefont
  {Y.}~\bibnamefont {Dumont}}, \bibinfo {author} {\bibfnamefont
  {C.}~\bibnamefont {Carrétéro}}, \bibinfo {author} {\bibfnamefont
  {M.}~\bibnamefont {Bibes}}, \ and\ \bibinfo {author} {\bibfnamefont
  {A.}~\bibnamefont {Barthélémy}},\ }\href {\doibase 10.1063/1.4861132}
  {\bibfield  {journal} {\bibinfo  {journal} {Applied Physics Letters}\
  }\textbf {\bibinfo {volume} {104}},\ \bibinfo {pages} {021920} (\bibinfo
  {year} {2014})}\BibitemShut {NoStop}%
\bibitem [{\citenamefont {Abbate}\ \emph {et~al.}(2002)\citenamefont {Abbate},
  \citenamefont {Zampieri}, \citenamefont {Prado}, \citenamefont {Caneiro},
  \citenamefont {Gonzalez-Calbet},\ and\ \citenamefont
  {Vallet-Regi}}]{Abbate2002}%
  \BibitemOpen
  \bibfield  {author} {\bibinfo {author} {\bibfnamefont {M.}~\bibnamefont
  {Abbate}}, \bibinfo {author} {\bibfnamefont {G.}~\bibnamefont {Zampieri}},
  \bibinfo {author} {\bibfnamefont {F.}~\bibnamefont {Prado}}, \bibinfo
  {author} {\bibfnamefont {A.}~\bibnamefont {Caneiro}}, \bibinfo {author}
  {\bibfnamefont {J.~M.}\ \bibnamefont {Gonzalez-Calbet}}, \ and\ \bibinfo
  {author} {\bibfnamefont {M.}~\bibnamefont {Vallet-Regi}},\ }\href {\doibase
  10.1103/PhysRevB.65.155101} {\bibfield  {journal} {\bibinfo  {journal} {Phys.
  Rev. B}\ }\textbf {\bibinfo {volume} {65}},\ \bibinfo {pages} {1} (\bibinfo
  {year} {2002})}\BibitemShut {NoStop}%
\bibitem [{\citenamefont {Green}\ \emph {et~al.}(2016)\citenamefont {Green},
  \citenamefont {Haverkort},\ and\ \citenamefont {Sawatzky}}]{Green2016}%
  \BibitemOpen
  \bibfield  {author} {\bibinfo {author} {\bibfnamefont {R.~J.}\ \bibnamefont
  {Green}}, \bibinfo {author} {\bibfnamefont {M.~W.}\ \bibnamefont
  {Haverkort}}, \ and\ \bibinfo {author} {\bibfnamefont {G.~A.}\ \bibnamefont
  {Sawatzky}},\ }\href {\doibase 10.1103/PhysRevB.94.195127} {\bibfield
  {journal} {\bibinfo  {journal} {Phys. Rev. B}\ }\textbf {\bibinfo {volume}
  {94}},\ \bibinfo {pages} {195127} (\bibinfo {year} {2016})}\BibitemShut
  {NoStop}%
\bibitem [{\citenamefont {Haverkort}\ \emph {et~al.}(2012)\citenamefont
  {Haverkort}, \citenamefont {Zwierzycki},\ and\ \citenamefont
  {Andersen}}]{Haverkort2012}%
  \BibitemOpen
  \bibfield  {author} {\bibinfo {author} {\bibfnamefont {M.~W.}\ \bibnamefont
  {Haverkort}}, \bibinfo {author} {\bibfnamefont {M.}~\bibnamefont
  {Zwierzycki}}, \ and\ \bibinfo {author} {\bibfnamefont {O.~K.}\ \bibnamefont
  {Andersen}},\ }\href {\doibase 10.1103/PhysRevB.85.165113} {\bibfield
  {journal} {\bibinfo  {journal} {Phys. Rev. B}\ }\textbf {\bibinfo {volume}
  {85}},\ \bibinfo {pages} {165113} (\bibinfo {year} {2012})}\BibitemShut
  {NoStop}%
\bibitem [{\citenamefont {Haverkort}(2016)}]{Haverkort2016}%
  \BibitemOpen
  \bibfield  {author} {\bibinfo {author} {\bibfnamefont {M.~W.}\ \bibnamefont
  {Haverkort}},\ }\href {\doibase 10.1088/1742-6596/712/1/012001} {\bibfield
  {journal} {\bibinfo  {journal} {Journal of Physics: Conference Series}\
  }\textbf {\bibinfo {volume} {712}},\ \bibinfo {pages} {012001} (\bibinfo
  {year} {2016})}\BibitemShut {NoStop}%
\bibitem [{\citenamefont {Eisebitt}\ \emph {et~al.}(1993)\citenamefont
  {Eisebitt}, \citenamefont {B{\"{o}}ske}, \citenamefont {Rubensson},\ and\
  \citenamefont {Eberhardt}}]{Eisebitt1993}%
  \BibitemOpen
  \bibfield  {author} {\bibinfo {author} {\bibfnamefont {S.}~\bibnamefont
  {Eisebitt}}, \bibinfo {author} {\bibfnamefont {T.}~\bibnamefont
  {B{\"{o}}ske}}, \bibinfo {author} {\bibfnamefont {J.-E.}\ \bibnamefont
  {Rubensson}}, \ and\ \bibinfo {author} {\bibfnamefont {W.}~\bibnamefont
  {Eberhardt}},\ }\href@noop {} {\bibfield  {journal} {\bibinfo  {journal}
  {Phys. Rev. B}\ }\textbf {\bibinfo {volume} {47}},\ \bibinfo {pages} {14103}
  (\bibinfo {year} {1993})}\BibitemShut {NoStop}%
\bibitem [{\citenamefont {Moriga}\ \emph {et~al.}(1994)\citenamefont {Moriga},
  \citenamefont {Usaka}, \citenamefont {Nakabayashi}, \citenamefont
  {Hirashima}, \citenamefont {Kohno}, \citenamefont {Kikkawa},\ and\
  \citenamefont {Kanamaru}}]{Moriga1994}%
  \BibitemOpen
  \bibfield  {author} {\bibinfo {author} {\bibfnamefont {T.}~\bibnamefont
  {Moriga}}, \bibinfo {author} {\bibfnamefont {O.}~\bibnamefont {Usaka}},
  \bibinfo {author} {\bibfnamefont {I.}~\bibnamefont {Nakabayashi}}, \bibinfo
  {author} {\bibfnamefont {Y.}~\bibnamefont {Hirashima}}, \bibinfo {author}
  {\bibfnamefont {T.}~\bibnamefont {Kohno}}, \bibinfo {author} {\bibfnamefont
  {S.}~\bibnamefont {Kikkawa}}, \ and\ \bibinfo {author} {\bibfnamefont
  {F.}~\bibnamefont {Kanamaru}},\ }\href {\doibase
  10.1016/0167-2738(94)90212-7} {\bibfield  {journal} {\bibinfo  {journal}
  {Solid State Ionics}\ }\textbf {\bibinfo {volume} {74}},\ \bibinfo {pages}
  {211} (\bibinfo {year} {1994})}\BibitemShut {NoStop}%
\bibitem [{\citenamefont {Alonso}\ \emph {et~al.}(1997)\citenamefont {Alonso},
  \citenamefont {Mart{\'{i}}nez-Lope}, \citenamefont
  {Garc{\'{i}}a-Mu{\~{n}}oz},\ and\ \citenamefont
  {Fern{\'{a}}ndez-D{\'{i}}az}}]{Alonso1997}%
  \BibitemOpen
  \bibfield  {author} {\bibinfo {author} {\bibfnamefont {J.~A.}\ \bibnamefont
  {Alonso}}, \bibinfo {author} {\bibfnamefont {M.~J.}\ \bibnamefont
  {Mart{\'{i}}nez-Lope}}, \bibinfo {author} {\bibfnamefont {J.~L.}\
  \bibnamefont {Garc{\'{i}}a-Mu{\~{n}}oz}}, \ and\ \bibinfo {author}
  {\bibfnamefont {M.~T.}\ \bibnamefont {Fern{\'{a}}ndez-D{\'{i}}az}},\ }\href
  {\doibase 10.1088/0953-8984/9/30/010} {\bibfield  {journal} {\bibinfo
  {journal} {Journal of Physics Condensed Matter}\ }\textbf {\bibinfo {volume}
  {9}},\ \bibinfo {pages} {6417} (\bibinfo {year} {1997})}\BibitemShut
  {NoStop}%
\bibitem [{\citenamefont {Trukhanov}\ \emph {et~al.}(2002)\citenamefont
  {Trukhanov}, \citenamefont {Troyanchuk}, \citenamefont {Pushkarev},\ and\
  \citenamefont {Szymczak}}]{Trukhanov2002}%
  \BibitemOpen
  \bibfield  {author} {\bibinfo {author} {\bibfnamefont {S.~V.}\ \bibnamefont
  {Trukhanov}}, \bibinfo {author} {\bibfnamefont {I.~O.}\ \bibnamefont
  {Troyanchuk}}, \bibinfo {author} {\bibfnamefont {N.~V.}\ \bibnamefont
  {Pushkarev}}, \ and\ \bibinfo {author} {\bibfnamefont {H.}~\bibnamefont
  {Szymczak}},\ }\href {\doibase 10.1134/1.1506439} {\bibfield  {journal}
  {\bibinfo  {journal} {Journal of Experimental and Theoretical Physics}\
  }\textbf {\bibinfo {volume} {95}},\ \bibinfo {pages} {308} (\bibinfo {year}
  {2002})}\BibitemShut {NoStop}%
\bibitem [{\citenamefont {Zhong}\ \emph {et~al.}(2003)\citenamefont {Zhong},
  \citenamefont {Jiang}, \citenamefont {Wu}, \citenamefont {Tang},
  \citenamefont {Chen},\ and\ \citenamefont {Du}}]{Zhong2003}%
  \BibitemOpen
  \bibfield  {author} {\bibinfo {author} {\bibfnamefont {W.}~\bibnamefont
  {Zhong}}, \bibinfo {author} {\bibfnamefont {H.~Y.}\ \bibnamefont {Jiang}},
  \bibinfo {author} {\bibfnamefont {X.~L.}\ \bibnamefont {Wu}}, \bibinfo
  {author} {\bibfnamefont {N.~J.}\ \bibnamefont {Tang}}, \bibinfo {author}
  {\bibfnamefont {W.}~\bibnamefont {Chen}}, \ and\ \bibinfo {author}
  {\bibfnamefont {Y.~W.}\ \bibnamefont {Du}},\ }\href {\doibase
  10.1088/0256-307X/20/5/343} {\bibfield  {journal} {\bibinfo  {journal}
  {Chinese Physics Letters}\ }\textbf {\bibinfo {volume} {20}},\ \bibinfo
  {pages} {742} (\bibinfo {year} {2003})}\BibitemShut {NoStop}%
\bibitem [{\citenamefont {Wang}\ \emph {et~al.}(2018)\citenamefont {Wang},
  \citenamefont {Rosenkranz}, \citenamefont {Rui}, \citenamefont {Zhang},
  \citenamefont {Ye}, \citenamefont {Zheng}, \citenamefont {Klie},
  \citenamefont {Mitchell},\ and\ \citenamefont {Phelan}}]{Wang2018}%
  \BibitemOpen
  \bibfield  {author} {\bibinfo {author} {\bibfnamefont {B.-X.}\ \bibnamefont
  {Wang}}, \bibinfo {author} {\bibfnamefont {S.}~\bibnamefont {Rosenkranz}},
  \bibinfo {author} {\bibfnamefont {X.}~\bibnamefont {Rui}}, \bibinfo {author}
  {\bibfnamefont {J.}~\bibnamefont {Zhang}}, \bibinfo {author} {\bibfnamefont
  {F.}~\bibnamefont {Ye}}, \bibinfo {author} {\bibfnamefont {H.}~\bibnamefont
  {Zheng}}, \bibinfo {author} {\bibfnamefont {R.~F.}\ \bibnamefont {Klie}},
  \bibinfo {author} {\bibfnamefont {J.~F.}\ \bibnamefont {Mitchell}}, \ and\
  \bibinfo {author} {\bibfnamefont {D.}~\bibnamefont {Phelan}},\ }\href
  {\doibase 10.1103/PhysRevMaterials.2.064404} {\bibfield  {journal} {\bibinfo
  {journal} {Phys. Rev. Materials}\ }\textbf {\bibinfo {volume} {2}},\ \bibinfo
  {pages} {064404} (\bibinfo {year} {2018})}\BibitemShut {NoStop}%
\bibitem [{\citenamefont {Jungwirth}\ \emph {et~al.}(2016)\citenamefont
  {Jungwirth}, \citenamefont {Marti}, \citenamefont {Wadley},\ and\
  \citenamefont {Wunderlich}}]{Jungwirth2016}%
  \BibitemOpen
  \bibfield  {author} {\bibinfo {author} {\bibfnamefont {T.}~\bibnamefont
  {Jungwirth}}, \bibinfo {author} {\bibfnamefont {X.}~\bibnamefont {Marti}},
  \bibinfo {author} {\bibfnamefont {P.}~\bibnamefont {Wadley}}, \ and\ \bibinfo
  {author} {\bibfnamefont {J.}~\bibnamefont {Wunderlich}},\ }\href {\doibase
  10.1038/nnano.2016.18} {\bibfield  {journal} {\bibinfo  {journal} {Nature
  Nanotechnology}\ }\textbf {\bibinfo {volume} {11}},\ \bibinfo {pages} {231}
  (\bibinfo {year} {2016})}\BibitemShut {NoStop}%
\end{thebibliography}%

\end{document}